\documentclass[a4paper,twocolumn,english,showpacs,aps]{revtex4-1}

\usepackage{graphicx}
\usepackage{dcolumn}
\usepackage{bm}
\usepackage{bmpsize}

\usepackage{natbib}

\begin{document}


\title{Rayleigh-B\'{e}nard convection with uniform vertical magnetic field}

\author{Arnab Basak}

\author{Rohit Raveendran}

\author{Krishna Kumar}
\email{kumar@phy.iitkgp.ernet.in}
\affiliation{Department of Physics,
 Indian Institute of Technology, Kharagpur-721302, India}

\date{\today}

\begin{abstract}
We present the results of direct numerical simulations of Rayleigh-B\'{e}nard convection in the presence of a uniform vertical magnetic field near instability onset. We have done simulations in  boxes with square as well as rectangular cross-sections in the horizontal plane. We have considered horizontal aspect ratio $\eta = L_y/L_x =1$ and $2$. The onset of the primary and secondary instabilities are strongly suppressed in the presence of the vertical magnetic field for $\eta =1$. The Nusselt number $\mathrm{Nu}$ scales with Rayleigh number $\mathrm{Ra}$ close to the primary instability as $[\mathrm{\{Ra - Ra_c (Q)\}/Ra_c (Q)} ]^{0.91}$, where $\mathrm{Ra_c (Q)}$ is the threshold for onset of stationary convection at a given value of the Chandrasekhar number $\mathrm{Q}$. $\mathrm{Nu}$ also scales with $\mathrm{Ra/Q}$ as $(\mathrm{Ra/Q})^{\mu}$. The exponent $\mu$ varies in the range $0.39 \le \mu \le 0.57$ for $\mathrm{Ra/Q} \ge 25$. The primary instability is stationary as predicted by Chandrasekhar. The secondary instability is temporally periodic for $\mathrm{Pr}=0.1$ but quasiperiodic for $\mathrm{Pr} = 0.025$ for moderate values of $\mathrm{Q}$. Convective patterns for higher values of $\mathrm{Ra}$ consist of periodic, quasiperiodic and chaotic wavy rolls above onset of the secondary instability for $\eta=1$.  In addition, stationary as well as time dependent cross-rolls are observed, as $\mathrm{Ra}$ is further raised.  The ratio $r_{\circ}/\mathrm{Pr}$ is independent of $\mathrm{Q}$ for smaller values of $\mathrm{Q}$. The delay in onset of the oscillatory instability is significantly reduced in a simulation box with $\eta =2$. We also observe inclined stationary rolls for smaller values of $\mathrm{Q}$ for $\eta =2$.
\end{abstract}

\maketitle


\section{\label{sec:Intro}Introduction}
Rayleigh-B\'{e}nard convection (RBC)~\cite{chandra}, where a thin horizontal layer of fluid is heated from below, is a topic of intense research. A particularly interesting variation of RBC is the case in which a low-Prandtl-number fluid is subjected to a magnetic field~\cite{chandra, nakagawa57, nakagawa59,  proctor_weiss_1982}. The system, also known as hydromagnetic convection, is relevant for several geophysical and astrophysical problems~\cite{glatzmaier_etal_1999, cattaneo_etal_2003, rucklidge06}. There have been extensive theoretical and numerical studies of RBC in fluids in the presence of an external magnetic field~\cite{chandra, nakagawa57, nakagawa59, proctor_weiss_1982, fauve_etal_1981, knobloch_etal_1981,  bc82, fauve_etal_1984, meneguzzi87, cb89,  houchens_2002, dawes_2007, podvigina_2010, pk_2012, cioni_etal_2000, ao_2001, bm_2002, yanagisawa_etal_2011}. They have addressed interesting issues such as pattern selection and instability~\cite{bc82, fauve_etal_1984, meneguzzi87, cb89, houchens_2002, dawes_2007, podvigina_2010, pk_2012}, heat transport~\cite{cioni_etal_2000, ao_2001, bm_2002}, and flow reversal~\cite{yanagisawa_etal_2011}. 

It is widely known that a uniform magnetic field tends to suppress the onset of convection, significantly reduces the convective heat transport across the fluid layer, and affects the primary as well as secondary instabilities~\cite{chandra, nakagawa57, nakagawa59, knobloch_etal_1981, bc82, cb89}. Experiments~\cite{cioni_etal_2000, ao_2001} show that the scaling exponent of the Nusselt number $\mathrm{Nu}$, which is a measure of the convective heat flux across the fluid layer, with the Rayleigh number $\mathrm{Ra}$ depends on the Chandrasekhar number $\mathrm{Q}$. The numerical simulations~\cite{bc82, cb89} have mainly focused on investigating  the stability of fluid patterns near primary instability. A systematic study of the convective flow structures and the scaling behavior even close to the onset of convection is lacking.

In this paper, we present the results of direct numerical simulations (DNS) for RBC in low-Prandtl number fluids ($\mathrm{Pr} \le 0.7$) in the presence of small uniform vertical magnetic field ($\mathrm{Q} \le 670$).  We have carried out the simulations in a three dimensional box ($L_x \times L_y \times L_z$).  We have considered boxes of square ($L_y = L_x$) and rectangular ($L_y = 2 L_x$) cross sections in the horizontal plane. Straight stationary rolls appear at the primary instability. The stationary straight rolls persist for higher values of $\mathrm{Ra}$ and moderate values of $\mathrm{Q}$. $\mathrm{Nu}$ scales with $\epsilon = [\mathrm{Ra} -\mathrm{Ra_c(Q)}]/\mathrm{Ra_c(Q)}$  as $\epsilon^{0.91}$ very close to the primary instability. The range of scaling expands with increase in $\mathrm{Q}$. The Nusselt number also scales with $\mathrm{Ra/Q}$ as $(\mathrm{Ra/Q})^\mu$ with $ 0.39 \le \mu \le 0.57$ for $\mathrm{Ra/Q} \ge 25$, which is consistent with the experimental results~\cite{cioni_etal_2000, ao_2001}. The secondary instability is always in the form of quasiperiodic or periodic waves along the roll axis in a square box. The secondary instability leads to periodic waves in a rectangular box. The patterns of straight stationary rolls, wavy rolls, quasi-periodic and chaotic wavy rolls and stationary oblique rolls are observed, as $\mathrm{Ra}$ is raised at a given value of $\mathrm{Q}$.

\section{\label{sec:Hydro}Hydromagnetic System}
We consider a thin layer of a low-Prandtl-number Boussinesq fluid of a reference density $\rho_0$, kinematic viscosity $\nu$, thermal diffusivity $\kappa$, magnetic diffusivity $\lambda$, and thermal expansion coefficient $\alpha$, which is confined between two horizontal surfaces separated by a distance $d$. The fluid layer is subjected to an adverse temperature gradient $\beta$ and a uniform vertical magnetic field $B_0$. A coordinate system is chosen such that the lower fluid surface is coincident with the $xy$ plane. The $z$ axis is positive along the vertically upward direction, which is also the direction of the applied magnetic field. The hydrodynamic equations are made dimensionless by measuring lengths in units of the fluid thickness $d$, time in units of viscous diffusion time $d^2/\nu$, temperature in units of $\nu\beta d/\kappa$, and the magnetic field in units of $B_0\nu/\lambda$. As the magnetic Prandtl number $\mathrm{Pm} = \nu/\lambda$ is usually of the order $10^{-6}$ or less for terrestrial fluids,  we  set  $\mathrm{Pm}$ equal to zero. The dynamics of RBC in the presence of a uniform vertical magnetic field  is then governed by the following set of dimensionless equations:

\begin{eqnarray}
&\partial_t\mathbf{v} +(\mathbf{v}\cdot\mathbf{\nabla})\mathbf{v} = -\nabla p + \nabla^2\mathbf{v} + \mathrm{Q}\partial_z\mathbf{b} + \mathrm{Ra}\theta\mathbf{e}_3,\label{ns}\\
&\mathrm{Pr}[\partial_t\theta + (\mathbf{v}\cdot\mathbf{\nabla})\theta] = \nabla^2\theta + {v}_3,\label{temp}\\
&\nabla^2\mathbf{b}=-\partial_z\mathbf{v},\label{mag}\\
&\mathbf{\nabla}\cdot\mathbf{v}=\mathbf{\nabla}\cdot\mathbf{b}=0,\label{other}
\end{eqnarray}
where $p (x, y, z, t)$ is the fluid pressure due to convection, $\mathbf{v}\thinspace(x,y,z,t)\equiv (v_1, v_2, v_3)$ is the fluid velocity, $\theta\thinspace(x, y, z, t)$ is the convective temperature field, $\mathbf{b}\thinspace(x,y,z,t)=(b_1, b_2, b_3)$ is the induced magnetic field due to convection,  and $\mathbf{e}_3$ is a unit vector directed in the positive direction of the $z$ axis. The induced magnetic field is slaved to the velocity in the limit of $\mathrm{Pm} \rightarrow 0$, and there is no independent dynamics for $\mathbf{b}$ field. The dynamics of RBC in the presence of a uniform vertical magnetic field is then governed by three dimensionless parameters: (i) Prandtl number $\mathrm{Pr}=\nu/\kappa$, which is the ratio of kinematic viscosity $\nu$ and thermal diffusivity $\kappa$, (ii) Rayleigh number ${\mathrm{Ra}} = \alpha\beta gd^4/(\nu\kappa)$, and (iii) Chandrasekhar number $\mathrm{Q}= B_0^2d^2/(4\pi\rho_0\nu\lambda)$. 

We assume idealized {\it stress-free} boundary conditions at the upper and lower surfaces, which may be a more useful approximation on a boundary between two liquids with a large difference in their viscosities. Almost stress-free boundary conditions were achieved in experiments by Goldstein and Graham~\cite{goldstein_graham_1969}. We also consider the bounding surfaces to be thermally conducting but electrically non-conducting. Teflon or some ethylene-vinyl-acetate (EVA) composite~\cite{lee_etal_2008} may serve this purpose in an experiment. The boundary conditions are thus given by
\begin{equation}
\frac{\partial v_1}{\partial z} = \frac{\partial v_2}{\partial z} = v_3 = \theta= b_1 = b_2 = \frac{\partial b_3}{\partial z} = 0 ~ \mbox{at z = 0, 1}.\label{bc}
\end{equation}
All fields are considered periodic in the horizontal plane. The magnetic field $\mathbf{b}^{p}$ inside the electrically non-conducting boundaries of magnetic permeability $\mu_p$ is determined by a scalar potential $\psi$ ($\mathbf{b}^p = \mathbf {\nabla}\psi$), which satisfies the Laplace equation $\nabla^2 \psi = 0$. Non-zero horizontal velocities of a liquid metal at the stress-free boundaries allow a surface current density $\mathbf{S}$ at the horizontal boundaries. As the horizontal magnetic fields in the fluid vanish at the boundaries, we have the condition $\mathbf{e}_3 \mathbf{\times} \mathbf{b}^{p}/\mu_p  = \mathbf{S} = S_1 \mathbf{e}_1 + S_2 \mathbf{e}_2$ to be satisfied at the boundaries. The surface currents are fixed by applying the condition $b_3^{p} = b_3$ at the boundaries.

\section{\label{sec:DNS}Direct Numerical Simulations and Results}
The components of the velocity field $\mathbf{v}\thinspace(x,y,z,t)$ and the magnetic field $\mathbf{b}\thinspace(x,y,z,t)$, the convective temperature field $\theta\thinspace(x,y,z,t)$ and the pressure $p\thinspace(x,y,z,t)$ are expanded consistent with the boundary conditions~(Eq.~\ref{bc}). The expansions of the fields are:
\begin{eqnarray}
{v_1} (x,y,z,t) &=& \sum_{l,m,n} U_{lmn}(t) e^{i(lk_xx+mk_yy)} \cos{(n\pi z)},\\
{v_2} (x,y,z,t) &=& \sum_{l,m,n} V_{lmn}(t) e^{i(lk_xx+mk_yy)} \cos{(n\pi z)},\\
{v_3} (x,y,z,t) &=& \sum_{l,m,n} W_{lmn}(t) e^{i(lk_xx+mk_yy)} \sin{(n\pi z)}, \\
{\theta} (x,y,z,t) &=& \sum_{l,m,n} \Theta_{lmn}(t) e^{i(lk_xx+mk_yy)} \sin{(n\pi z)}, \\
{p} (x,y,z,t) &=& \sum_{l,m,n} P_{lmn}(t) e^{i(lk_xx+mk_yy)} \cos{(n\pi z)}.
\label{eq.DNS_expansion}
\end{eqnarray}
\noindent where $\mathbf{k} = k_x \mathbf{e}_1 + k_y \mathbf{e}_2$ is the wave vector of convective fields in the horizontal plane. The integers $l,m,n$ can take values consistent with the equation of continuity. The expansions of the magnetic fields are determined by Eq.~\ref{mag}. The scalar potentials $\psi |_{z\le 0}$ and $\psi |_{z\ge 1}$ in the lower and upper boundaries, respectively, are:
\begin{equation}
\psi |_{z\le 0} (x,y,z,t) = \sum_{l,m,n}  \Psi_{lmn} (x,y,t) e^{\gamma z}
\end{equation}
\begin{equation}
\psi |_{z\ge 1} (x,y,z,t) = \sum_{l,m,n} (-1)^{n+1}\Psi_{lmn} (x,y,t) e^{\gamma (1-z)},
\end{equation}
where
\begin{equation}
\Psi_{lmn} (x,y,t) = \frac{n\pi W_{lmn}(t) e^{ik_c(lx + my)}}{\gamma (\gamma^2 + n^2\pi^2)}
\end{equation}
and
\begin{equation}
\gamma = k_c \sqrt{(l^2 + m^2)}.
\end{equation}
The surface current densities on the boundaries ($z=0, 1$) can be computed from the following equation:
\begin{equation}
\mathbf{S} (z = 0, 1) = \mathbf{e}_3 \mathbf{\times}\mathbf{\nabla}\psi (z = 0, 1)/\mu_p
\end{equation} 

The critical Rayleigh number $\mathrm{Ra}_c (\mathrm{Q})$ and the critical wave number $k_c (\mathrm{Q})$ for the stationary convection with free-slip boundary conditions are given by,  
\begin{eqnarray}
\mathrm{Ra}_c (\mathrm{Q}) &=& \frac{\pi^2+k_c^2}{k_c^2}[(\pi^2+k_c^2)^2+\pi^2 \mathrm{Q}],\\
k_c (\mathrm{Q}) &=& \pi \sqrt{a_{+} + a_{-} - (1/2)},
\end{eqnarray}
\begin{equation}
a_{\pm} = \left( \frac{1}{4}\left\lbrace \frac{1}{2}+\frac{\mathrm{Q}}{\pi^2}\pm \left[ \left( \frac{1}{2} +\frac{\mathrm{Q}}{\pi^2}\right)^2 -\frac{1}{4}\right]^\frac{1}{2} \right\rbrace \right)^\frac{1}{3}.
\end{equation}

The full hydromagnetic system [Eqs.~(\ref{ns})-(\ref{other})] with the boundary conditions~(Eq.~\ref{bc}) has been integrated using a pseudo-spectral method in a three-dimensional simulation box ($L_x = 2\pi/k_x \times L_y = 2\pi/k_y \times L_z =1$). Simulation boxes with square ($\eta = L_y/ L_x = k_x/k_y = 1$) as well as rectangular ($\eta =2$) cross sections in the horizontal plane have been considered.  We have taken $k_x = k_y = k_{c} (\mathrm{Q})$ for the case of square cross-section, while  $k_x = k_{c} (\mathrm{Q})$ and $k_y = k_{c} (\mathrm{Q})/2$ for the case of rectangular cross-section. This allows us to investigate the interaction of two dimensional (2D) stationary straight rolls (Rolls) with a new set of rolls aligned perpendicular to the old set. The new rolls have the same wavelength as that of the old set for $\eta =1$ and double the wavelength of the old set for $\eta =2$. A spatial resolution of $64\times 64\times 64$ grids is used for simulations.  Time integration is carried out using a standard RK4 method with a maximum step size of 0.001. We vary the reduced Rayleigh number $r = \mathrm{Ra}/\mathrm{Ra}_c (\mathrm{Q})$ in small steps for a fixed value $\mathrm{Q}$ in our simulations. Final values of all fields for a given value of $r$ are used as initial conditions for the next higher value of $r$. Results are also verified for different values of $r$ starting with random initial conditions. They yield the same results. $\mathrm{Q}$ is varied from $5$ to $670$. 

\begin{figure}[ht]
\begin{center}
\resizebox{0.5\textwidth}{!}{%
  \includegraphics{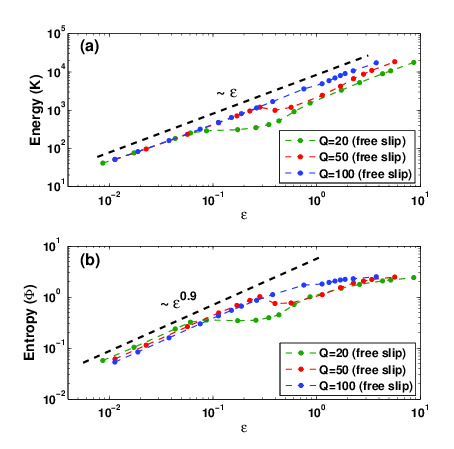}}
\caption{\label{fig:K_Phi} (Color online) Variation of (a) kinetic energy $K$ and (b) convective entropy $\Phi$  with $\epsilon=[\mathrm{Ra - Ra_c \thinspace(Q)}]/ \mathrm{Ra_c \thinspace(Q)}$ for $\mathrm{Pr}=0.1$ in a square box  [$k_x=k_y=k_c\thinspace(\mathrm{Q}) (\eta =1$)] shown in different colors for different values of $\mathrm{Q}$, as computed from DNS. Color (gray) dots are for different values of $\mathrm{Q}$. The dashed lines show the variation of (a) $K$ and (b) $\Phi$ with $\epsilon$ for the stationary convection.}
\end{center}
\end{figure}

\begin{figure*}[ht]
\begin{center}
\resizebox{1.0\textwidth}{!}{%
  \includegraphics{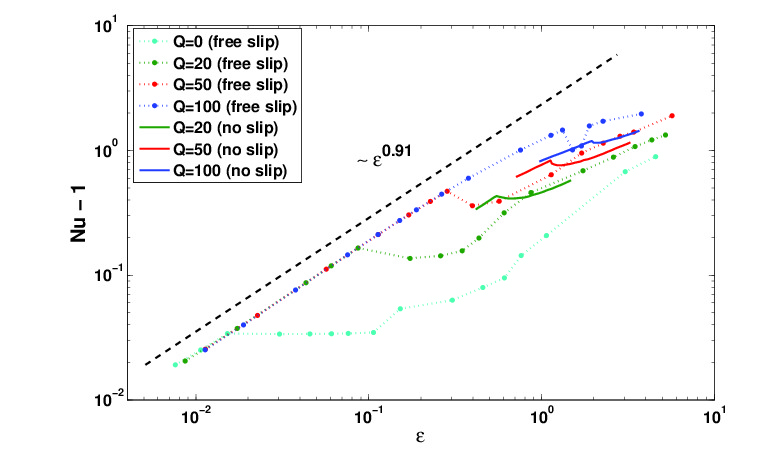}}
\caption{\label{fig:nusselt} (Color online) Plot of convective heat flux $(\mathrm{Nu}-1)$ as a function of $\epsilon=[\mathrm{Ra - Ra_c \thinspace(Q)}]/ \mathrm{Ra_c \thinspace(Q)}$ for $\mathrm{Pr}=0.1$ and $k_x=k_y=k_c\thinspace(\mathrm{Q}) (\eta =1)$ shown for four different $\mathrm{Q}$ values, as computed from DNS, for free slip boundaries (dotted lines with points). They are compared with the numerical results  (Clever and Busse~\cite{cb89}) for no slip boundaries (solid lines). The dashed line is parallel to the linear region for stationary rolls at onset for all four $\mathrm{Q}$ values. $(\mathrm{Nu}-1)$ scales as $\epsilon^{0.91}$ for stationary rolls near onset.}
\end{center}
\end{figure*}

\begin{figure}[ht]
\begin{center}
\resizebox{0.5\textwidth}{!}{%
  \includegraphics{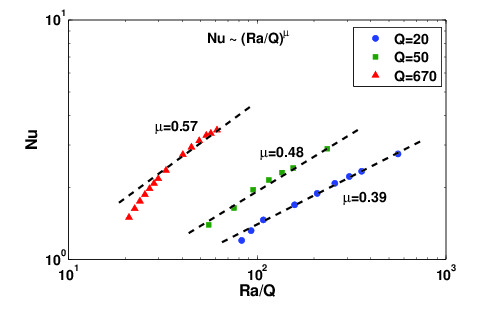}
}
\caption{\label{fig:nu_ra} (Color online) Variation of Nusselt number $\mathrm{Nu}$ with $\mathrm{Ra/Q}$ for $\mathrm{Pr}=0.1$ (square box) for three different values of $\mathrm{Q}$ (for $\mathrm{Ra/Q}>25$), as computed from DNS. The blue, green and red dashed lines are the best linear fits for  $\mathrm{Q} = 20, 50$ and $670$, respectively. $\mathrm{Nu}$ scales as $(\mathrm{Ra/Q})^{\mu}$, where $\mu$ values are $0.39$, $0.48$ and $0.57$ for $\mathrm{Q}=20$, $50$ and $670$ respectively.}
\end{center}
\end{figure}

\subsection{Scaling behavior near onset of convection}

We now present the scaling properties of global quantities like the time averaged kinetic energy per unit mass $K = \frac{1}{2} \int v^2 dV$ and ``convective entropy"  defined as $\Phi = \frac{1}{2} \int \theta^2 dV$ with $\epsilon=\mathrm{[Ra - Ra_c\thinspace(Q)] /Ra_c\thinspace(Q)}$. The convection sets in as stationary straight rolls at the primary instability. The vertical magnetic field delays the onset of convection, which is well known since the prediction of Chandrasekhar~\cite{chandra}. It is clearly evident that the secondary instability, which is oscillatory, is strongly inhibited due to the presence of a vertical magnetic field.  Figures~\ref{fig:K_Phi}(a) and (b) show the variation of $K$ and  $\Phi$ with $\epsilon$ for $\mathrm{Pr} = 0.1$ and three different values of $\mathrm{Q}$. The kinetic energy is proportional to  $\epsilon$ in the stationary convection regime. The deviation from this behavior is observed when there is a transition from stationary to oscillatory convection. Away from the onset of secondary (oscillatory) instability, $K$ is again approximately proportional to $\epsilon$.  The average flow speed scales with $\epsilon$ as $\epsilon^{1/2}$ near onset. However, the convective entropy scales with $\epsilon$ as $\epsilon^{0.9}$. The average convective temperature field therefore scales as $\epsilon^{0.45}$. The scaling of the average convective temperature field is different from that of the average speed. As soon as convection begins, there is advection of the temperature field and a thermal current is established in the vertical direction. A part of the thermal energy injected externally is used in maintaining the thermal flux. This is also reflected by two types of possible temperature modes $\Theta_{lmn}$ ($l$ or $m \neq 0$ and $n \ge 1$) and $\Theta_{00n} (n \ge 1)$. The modes $\Theta_{lmn}$ have zero horizontal average and they scale like velocity modes $W_{lmn}$. The temperature modes $\Theta_{00n}$ do not have a counterpart in the expansion for the vertical velocity as there is no net momentum flux in the vertical direction. Only in the limit of $\mathrm{Pr} \rightarrow 0$ would the convective thermal flux vanish and the temperature field follow the vertical velocity field.

The Nusselt number defined as $\mathrm{Nu} = 1 + \mathrm{Pr}^2 <v_3 \theta>_{xyz}$ is a measure of the heat flux across the fluid layer. The symbol $<.>_{xyz}$ stands for the spatial average over the simulation box.
The variation of the time averaged value of ($\mathrm{Nu}-1$) is plotted as a function of $\epsilon$ in Fig.~\ref{fig:nusselt} for $\mathrm{Pr} = 0.1$ at different values of $\mathrm{Q}$. The dotted curves in different colors (shades of gray) are the results of DNS for different values of $\mathrm{Q}$. The scaling of convective heat flux with $\epsilon$ near onset is given as: $\mathrm{Nu}-1 \sim \epsilon^{0.91}$.  The product $v_3 \theta$ scales as $\epsilon^{0.95}$, if we take  $v_3 \sim \sqrt{K} \sim \epsilon^{0.5}$ and $\theta \sim \sqrt{\Phi} \sim \epsilon^{0.45}$. This estimate is not accurate. The convective heat flux is proportional to $<v_3 \theta>_{xyz} = <W_{lmn}^{\star} \Theta_{lmn} + W_{lmn} \Theta_{lmn}^{\star}>_{xyz}$, and therefore all temperature modes do not contribute to the heat flux.  This feature in the scaling of heat transport should be observable in other variations of a Rayleigh-B\'{e}nard system. The scaling exponent for variation of the Nusselt number with $\epsilon$ may depend on the details of a particular system but it is expected to be less than unity. The increase of the Nusselt number with $\epsilon$ is stopped at the onset of oscillatory instability. It first decreases, reaches a minimum, and then begins to increases once again, as $\epsilon$ is raised in small steps. Solid lines in Fig.~\ref{fig:nusselt} display the variation of $\mathrm{Nu}$ with $\epsilon$ from numerical data obtained by Clever and Busse~\cite{cb89} for no-slip horizontal boundaries. The slope of the heat flux across the fluid layer during stationary convection and much after the onset of oscillatory convection with free-slip and no-slip boundary conditions shows qualitatively similar behavior, although the actual value of $\mathrm{Nu}$ with free-slip  is higher than its value with no-slip velocity boundary conditions. The convective heat transport for stationary convection scales with $\epsilon$ as $\epsilon^{0.91}$ near the primary instability for moderate values of $\mathrm{Q}$. For higher values of $\mathrm{Q}$ ($> 100$), the scaling exponent decreases at relatively higher values of $\epsilon$. The scaling law is broken at onset of the oscillatory convection. However, the $\mathrm{Nu} - \epsilon$ curves with free-slip and no-slip conditions show similar slopes for higher values of $\epsilon$. The slope of any $\mathrm{Nu} - \epsilon$ curve for time-dependent convection is smaller than its value for stationary convection. 

The scaling exponent of $\mathrm{Nu}$ is slightly less than unity for smaller values of $\mathrm{Ra}$, which is expected even in the absence of magnetic field. The scaling exponent is likely to be dominated by the buoyancy force at much higher values of $\mathrm{Ra}$, and is therefore expected to be independent of $\mathrm{Q}$ for $\mathrm{Ra} \gg \mathrm{Q}$ (fully developed turbulent regime). One expects an intermediate regime where the Lorentz force as well as the force of buoyancy together may decide the scaling exponent, if a scaling behavior is possible. We have therefore tried to probe the possibility of a scaling regime by plotting $\mathrm{Nu}$ as a function of $\mathrm{Ra/Q}$. Figure~\ref{fig:nu_ra} shows the variation of $\mathrm{Nu}$ as a function of $\mathrm{Ra/Q}$ for $\mathrm{Pr}=0.1$. The  points shown as  blue (black) circles, green (gray) squares and red (light gray) triangles are data points computed from DNS for  $\mathrm{Q} = 20$, $50$ and $670$, respectively. They correspond to $\mathrm{Ra/Q}$ values in a range between $40$ and $600$. The dashed lines, which are the best linear fits of the data obtained from DNS, show that $\mathrm{Nu}$ scales with $\mathrm{Ra/Q}$ as $(\mathrm{Ra/Q})^\mu$. The scaling exponent $\mu$ is found to vary with $\mathrm{Q}$. Values of the exponent $\mu$ are found to be   $0.39$, $0.48$ and $0.57$ for $\mathrm{Q}=20$, $50$ and $670$, respectively. It is interesting to note that the value of $\mu$ was found to be equal to $0.50 \pm 0.03$ for $\mathrm{Q}=670$ in experiments by Aurnou and Olson~\cite{ao_2001}. Its value was found to be $0.43$ for $\mathrm{Ra/Q} \approx 50$ in experiments by Cioni et al.~\cite{cioni_etal_2000}. The latter case was a regime of soft convective turbulence  with reduced Rayleigh number $r \approx 100$.  The value of $\mu$ computed in DNS are in qualitative agreement with those observed in experiments. DNS shows a decrease in $\mu$ with an increase in $\mathrm{Ra/Q}$ value.

\begin{table*}[ht]
\caption{\label{tab:patt_sq} Convective patterns in a square simulation box ($\eta = 1$) computed from DNS for (i) $\mathrm{Pr}=0.025$ and (ii) $\mathrm{Pr}=0.1$ for different values of $r$. Patterns observed are: 2D stationary rolls (Rolls), periodic wavy rolls (WR), quasiperiodic wavy rolls (QWR), stationary cross rolls (CR), quasiperiodic cross rolls (QPCR) and chaotic cross rolls (CCR).}
\begin{ruledtabular}
\begin{tabular}{ccccccccccc}
 &\multicolumn{5}{c}{$\mathrm{Pr}=0.025$}&\multicolumn{5}{c}{$\mathrm{Pr}=0.1$}\\
 $\mathrm{Q}$&$r=1.01$&$r=1.05$&$r=1.1$&$r=1.2$&$r=1.4$
&$r=1.01$&$r=1.05$&$r=1.1$&$r=1.2$&$r=1.4$\\ \hline
 5&Rolls&QWR&QWR&CCR&CR &Rolls&WR&WR&QPCR&CR \\
 10&Rolls&QWR&QWR&CCR&CR &Rolls&WR&WR&WR&CCR\\
 20&Rolls&QWR&QWR&CCR&CCR &Rolls&Rolls&WR&WR&QPCR\\
 30&Rolls&QWR&WR&WR&CCR &Rolls&Rolls&Rolls&WR&WR\\
 40&Rolls&QWR&CWR&WR&CCR &Rolls&Rolls&Rolls&WR&WR\\
 50&Rolls&QWR&CWR&WR&CCR &Rolls&Rolls&Rolls&Rolls&WR\\
 60&Rolls&Rolls&QWR&WR&WR &Rolls&Rolls&Rolls&Rolls&Rolls\\
 70&Rolls&Rolls&Rolls&QWR&WR &Rolls&Rolls&Rolls&Rolls&Rolls\\
 80&Rolls&Rolls&Rolls&WR&WR &Rolls&Rolls&Rolls&Rolls&Rolls\\
 90&Rolls&Rolls&Rolls&WR&WR &Rolls&Rolls&Rolls&Rolls&Rolls\\
 100&Rolls&Rolls&Rolls&WR&CWR &Rolls&Rolls&Rolls&Rolls&Rolls\\
 120&Rolls&Rolls&Rolls&Rolls&WR &Rolls&Rolls&Rolls&Rolls&Rolls\\
\end{tabular}
\end{ruledtabular}
\end{table*}

\begin{figure}[ht]
\begin{center}
\resizebox{0.45\textwidth}{!}{%
  \includegraphics{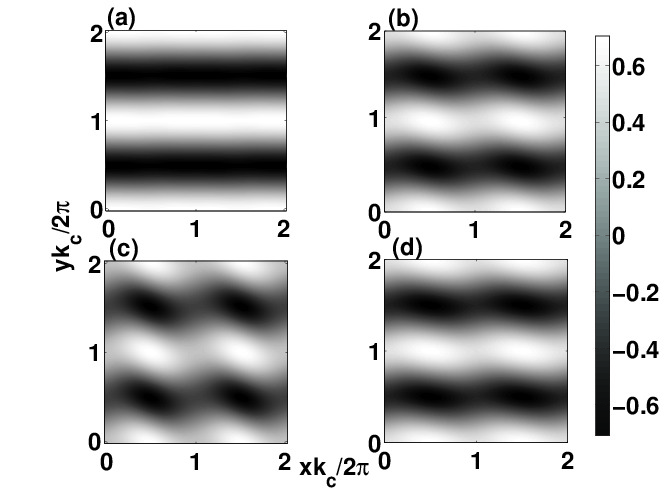}
}
\caption{\label{fig:quasi_cont} Contour plots of the convective temperature field at $z=0.5$ close to onset of convection ($r=1.05$) in a simulation box with square cross section ($\eta = 1$) for $\mathrm{Pr}=0.025$  and $\mathrm{Q}=50$ [$k_\mathrm{c} (\mathrm{Q}=50) = 3.270$] showing temporally quasiperiodic wavy rolls (QWR). Straight rolls are observed whenever the Fourier mode $W_{111}$ becomes zero.}
\end{center}
\end{figure}

\begin{figure}[h]
\begin{center}
\resizebox{0.5\textwidth}{!}{%
  \includegraphics{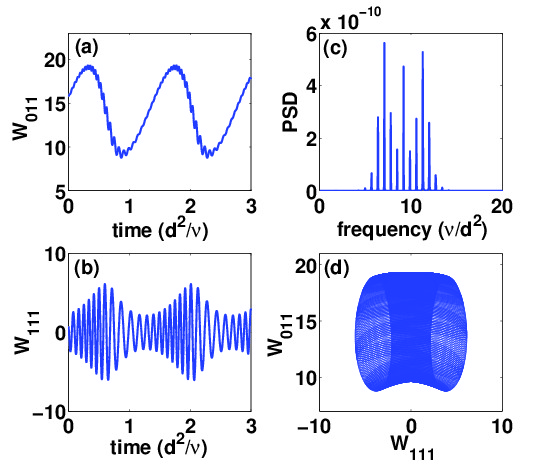}
}
\caption{\label{fig:quasi} (Color online) Properties of quasiperiodic wavy rolls (QWR) corresponding to parameters given in Fig.~\ref{fig:quasi_cont}. Temporal  variations of the two largest Fourier modes: (a) $W_{011}$ and (b) $W_{111}$, (c) power spectral density (PSD) of the Fourier mode $W_{111}$ and (d) phase portrait in the $W_{111}-W_{011}$ plane show quasiperiodic nature of the wavy patterns.}
\end{center}
\end{figure}

\subsection{Fluid patterns in a square simulation box}
Abrupt changes in the $\mathrm{Nu} - \epsilon$ curves (see Fig.~\ref{fig:nusselt}) for a fixed value $\mathrm{Q}$ indicate bifurcations in the convective flow structures. We have computed the fluid patterns from DNS. Table~\ref{tab:patt_sq} enlists the convective patterns for a square box $\eta = 1$ for $\mathrm{Pr}=0.025$ and $\mathrm{Pr}=0.1$ close to onset. Thermal convection appears as two-dimensional (2D) stationary convection (Rolls) at the primary instability. For larger values of $\mathrm{Q}$ ($\ge 60$), the 2D rolls remain stable even at $r = 1.4$ for $\mathrm{Pr} =0.1$. The secondary instability is strongly delayed in the presence of a vertical magnetic field. Time dependent convection appears at the secondary instability. Fluid patterns show quasiperiodic wavy rolls for $\mathrm{Pr} = 0.025$ and periodic wavy rolls for $\mathrm{Pr} = 0.1$. As the reduced Rayleigh number $r$ is raised for a fixed value of $\mathrm{Q}$,  three dimensional (3D) convective patterns consisting of periodic wavy rolls (WR), quasiperiodic wavy rolls (QWR), and chaotic wavy rolls (CWR) are observed.  Further increase in $r$ leads to stationary cross rolls (CR), quasiperiodic cross-rolls (QPCR) and chaotic cross-rolls (CCR). Rolls become unstable, if $r$ is increased  for fixed values of $\mathrm{Pr}$ and $\mathrm{Q}$. However, increase in $\mathrm{Q}$ for a fixed values of $\mathrm{Pr}$ and $r$ always leads to 2D stationary rolls for sufficiently large vales of $\mathrm{Q}$. 

\begin{figure}[ht]
\begin{center}
\resizebox{0.45\textwidth}{!}{%
  \includegraphics{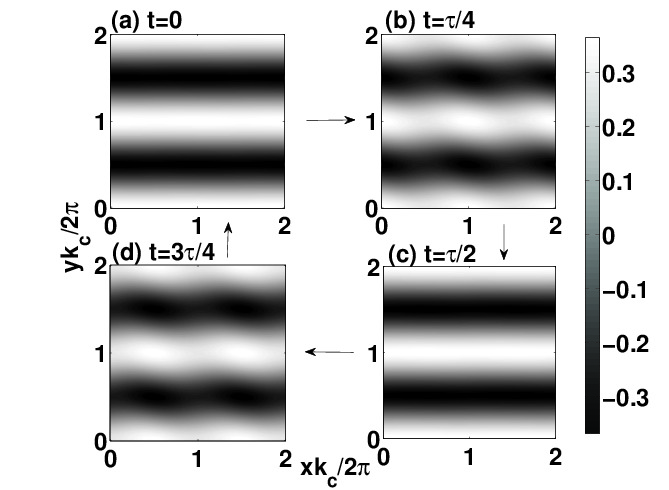}
}
\caption{\label{fig:wavy_cont} Contour plots of the temperature field at mid plane ($z=0.5$) in a simulation box with square cross section [$L_x = L_y = 2\pi/k_\mathrm{c}\thinspace(\mathrm{Q})$] for $\mathrm{Pr}=0.1$, $r=1.05$. Periodic wavy rolls (WR) for $\mathrm{Q} = 10$ [$k_\mathrm{c} (\mathrm{Q}=10) = 2.589$] with time period $\tau = 0.262$ (in viscous time units) at four instants: (a) $t = 0$, (b) $t = \tau/4$, (b) $t = \tau/2$ and (b) $t = 3\tau/4$.}
\end{center}
\end{figure}

\begin{figure}[ht]
\begin{center}
\resizebox{0.5\textwidth}{!}{%
  \includegraphics{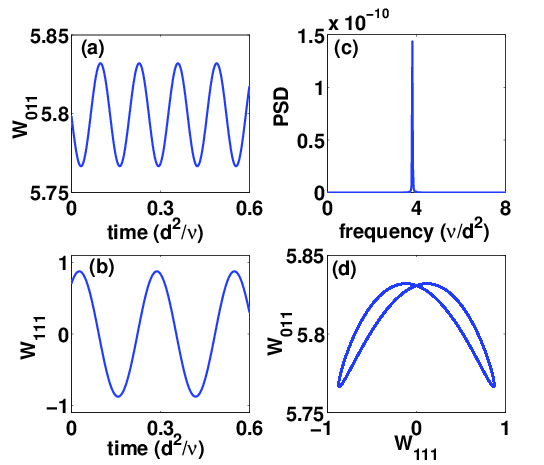}
}
\caption{\label{fig:wavy} (Color online) Properties of periodic wavy rolls (WR) in a simulation box with square cross section corresponding to the parameters given in Fig.~\ref{fig:wavy_cont}. The temporal variations of the two largest Fourier modes (a) $W_{011}$ and (b) $W_{111}$, (c) the power spectral density (PSD) of the mode $W_{111}$, and (d) the phase portrait in the $W_{111}-W_{011}$ plane show a set of periodic wavy rolls.}
\end{center}
\end{figure}

\begin{figure}[ht]
\begin{center}
\resizebox{0.45\textwidth}{!}{%
  \includegraphics{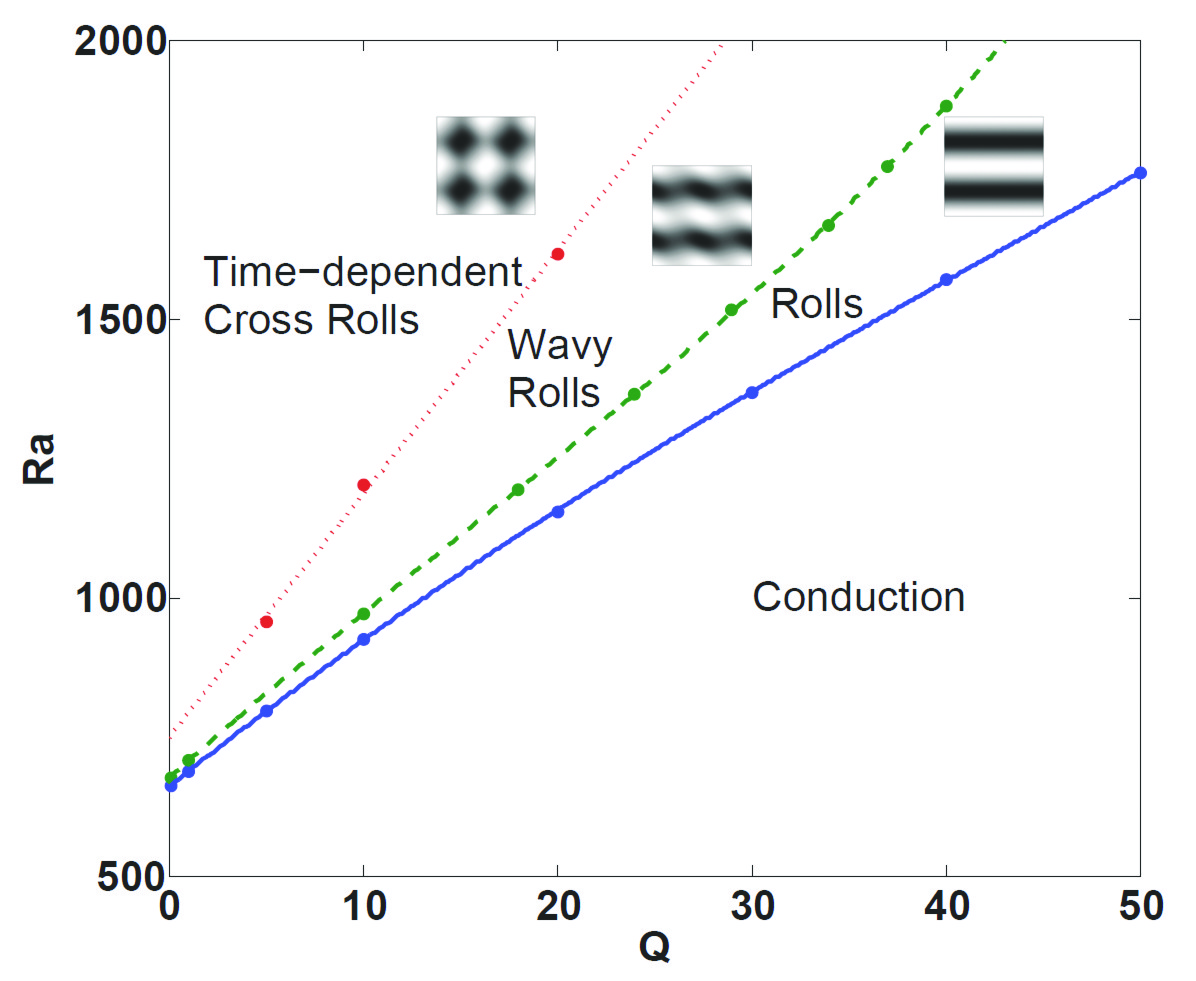}
}
\caption{\label{fig:regions} (Color online) Plot of $\mathrm{Ra}$ versus $\mathrm{Q}$ showing various regions for $\mathrm{Pr}=0.1$ and $k_x=k_y=k_\mathrm{c}\thinspace(\mathrm{Q})$. The onsets of primary as well as secondary instabilities are delayed by magnetic field. Convection at onset is stationary (Rolls) and becomes oscillatory (WR) after $\mathrm{Ra}$ is increased above $\mathrm{Ra_o\thinspace(Q, Pr)}$, the threshold for secondary (oscillatory) instability. Above the region of wavy rolls, we observe time-dependent cross-rolls. The dashed lines show polynomial fits for the DNS data (points).}
\end{center}
\end{figure}

\begin{figure}[ht]
\begin{center}
\resizebox{0.5\textwidth}{!}{%
  \includegraphics{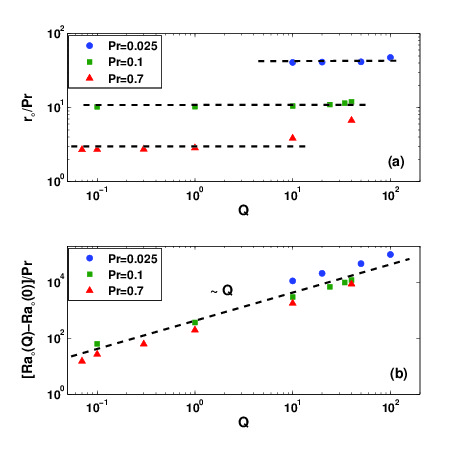}
}
\caption{\label{fig:scalings} (Color online) Scaling of oscillatory threshold with $\mathrm{Q}$. (a) The combination $r_\mathrm{\circ}\mathrm{\thinspace(Q, Pr)/Pr}$ is independent of $\mathrm{Q}$ for smaller values of $\mathrm{Q}$. Circles, squares and triangles are data points for $\mathrm{Pr}=0.025$, $\mathrm{Pr}=0.1$ and $\mathrm{Pr}=0.7$, respectively. (b) The quantity $\mathrm{[Ra_{\circ} (Q) - Ra_{\circ} (Q=0)]/Pr}$ varies linearly with $\mathrm{Q}$ for different values of $\mathrm{Pr}$.}
\end{center}
\end{figure}

Figure~\ref{fig:quasi_cont} shows the contour plots of the convective temperature field at the mid-plane ($z=1/2$) of a square simulation box for $\mathrm{Pr}=0.025$, $\mathrm{Q}=50$ and $r=1.05$ at four time instants at equal interval of one-fourth of the faster time period. The time dependent patterns alternate between straight and wavy rolls.  Temporal variations of the two largest Fourier modes $W_{011}$ and $W_{111}$  corresponding to these patterns are shown in Figs.~\ref{fig:quasi}(a) and (b), respectively.  The mode $W_{101}$ is not excited in this case. The Fourier mode $W_{011}$ oscillates with a non-zero mean, while the mode $W_{111}$ oscillates with zero mean. Both the Fourier modes show amplitude modulation. The power spectral density (PSD) of the mode $W_{111}$ [Fig.~\ref{fig:quasi}(c)] and the phase portrait in the $W_{111}-W_{011}$ plane suggests temporally quasiperiodic nature of the patterns [Fig.~\ref{fig:quasi}(d)]. The secondary instability is always in the form of quasiperiodic wavy rolls (QWR) for $\mathrm{Pr} = 0.025$.

Figure~\ref{fig:wavy_cont} shows the mid-plane contour plots for a temperature field for $\mathrm{Pr}=0.1$, $r=1.05$ and $\mathrm{Q}=10$ at equal time interval equal to one-fourth of the period of oscillation $\tau$.  Temporal variations of the two largest Fourier modes $W_{011}$ and $W_{111}$ are shown in Figs.~\ref{fig:wavy}(a) and (b), respectively. The mode $W_{011}$ again oscillates with a non-zero mean, while the mode $W_{111}$ oscillates with zero mean. The period of the wavy mode $W_{111}$ is equal to  double the period of 2D roll mode $W_{011}$. The power spectral density (PSD) of the mode $W_{111}$ [Fig.~\ref{fig:wavy}(c)] and the phase portrait in the $W_{111}-W_{011}$ plane [Fig.~\ref{fig:wavy}(d)] also confirm the periodic waves along the roll axis. The amount of waviness in the rolls at any given instant depends on the value of $W_{111}$ at that  instant. 

Figure~\ref{fig:regions} shows the regions of the parameter space  (the $\mathrm{Ra - Q}$ plane) for $\mathrm{Pr}=0.1$ and $\eta = 1$ having different convective patterns, as obtained from DNS. The region of the $\mathrm{Ra - Q}$ plane below the solid curve represents the conduction state.  The solid curve is the threshold value of the Rayleigh number $\mathrm{Ra_c}\thinspace(\mathrm{Q})$ for the appearance of thermal convection, as obtained by Chandrasekhar~\cite{chandra}. The points on this curve are computed from DNS. They are in complete agreement.  As we have chosen $\mathrm{Pm} \rightarrow 0$, the condition for stationary convection at onset ($\mathrm{Pr} > \mathrm{Pm}$) always holds. The onset of convection is always stationary, and 2D straight rolls are the primary convective patterns.  Straight rolls become unstable due to oscillatory instability. The dashed curve represents the threshold for secondary instability $\mathrm{Ra}_{\circ}$, which is oscillatory in this case.  Periodic wavy rolls (WR) are observed in the region of the $\mathrm{Ra - Q}$ plane bounded by the dashed and the dotted curves. In the region above the dotted curve, we observe a time-dependent competition of two sets of rolls in mutually perpendicular directions. Both sets of rolls oscillate around a finite mean.  Consequently, the fluid patterns consist of time dependent cross-rolls. The presence of a vertical magnetic field  strongly delays not only the primary instability but also the secondary and tertiary instabilities. The threshold for the oscillatory instability $\mathrm{Ra}_{\circ}$ increases monotonically with $\mathrm{Q}$ which shows the inhibition of the oscillatory instability, in agreement with the observations of Clever and Busse \cite{cb89}. 

The threshold for oscillatory (secondary) instability $\mathrm{Ra_{\circ} (Q, Pr)}$ also shows scaling behavior with $\mathrm{Q}$. The variation of  $r_{\mathrm{\circ}}\mathrm{\thinspace(Q, Pr)/Pr}$, where $r_{\mathrm{\circ}} = \mathrm{[Ra_{\circ} (Q, Pr)/ Ra_c (Q)]}$, with $\mathrm{Q}$ is plotted in Fig.~\ref{fig:scalings}(a) for different values of $\mathrm{Pr}$. The quantity $r_{\mathrm{\circ}} \mathrm{\thinspace(Q, Pr)/Pr}$ is found to be independent of $\mathrm{Q}$ for lower values of $\mathrm{Q}$. This behavior is observed in a wider range of $\mathrm{Q}$ for smaller values of $\mathrm{Pr}$. Figure~\ref{fig:scalings}(b) shows the variation of $[\mathrm{Ra_{\circ} (Q)-Ra_{\circ} (Q = 0)]/Pr}$ with $\mathrm{Q}$ for different values of $\mathrm{Pr}$. The quantity $[\mathrm{Ra_{\circ} (Q)-Ra_{\circ} (Q=0)]/Pr}$ is found to be proportional to $\mathrm{Q}$. This type of scaling was also observed in a low-dimensional model of thermal convection in the presence of a uniform horizontal magnetic field~\cite{pk_2012}. 

\subsection{Fluid patterns in a rectangular simulation box}
It is known that the straight rolls are unstable to long wavelength perturbations in low-Prandtl-number fluids in the absence of any magnetic field. The fluid patterns in a rectangular simulation box ($\eta =2$) are therefore likely to show interesting behavior. Table~\ref{tab:patt_rect} lists the possible convective patterns in low-Prandtl-number fluids ($\mathrm{Pr} = 0.025$ and $\mathrm{Pr} = 0.1$) computed from DNS in a rectangular simulation box. Straight stationary rolls appear at the onset of convection as observed in a square simulation box with $\eta = 2$. These straight rolls become unstable and wavy rolls are excited at the secondary instability. Standing waves are generated along the roll axis.

\begin{table}[h]
\caption{\label{tab:patt_rect} Convective patterns in a rectangular simulation box ($\eta =2$) computed from DNS for (i) $\mathrm{Pr}=0.025$ and (ii) $\mathrm{Pr}=0.1$ for different values of $r$. Patterns observed are: 2D stationary rolls (Rolls), stationary inclined rolls (IR), periodic wavy rolls (WR), quasiperiodic wavy rolls (QWR), and chaotic wavy rolls (CWR).}
\begin{ruledtabular}
\begin{tabular}{ccccccc}
 &\multicolumn{3}{c}{$\mathrm{Pr}=0.025$}&\multicolumn{3}{c}{$\mathrm{Pr}=0.1$}\\
 $\mathrm{Q}$&$r=1.05$&$r=1.1$&$r=1.2$
&$r=1.05$&$r=1.1$&$r=1.2$\\ \hline
 5&WR&CWR&CWR &WR&QWR&IR \\
 10&WR&CWR&CWR &WR&WR&IR\\
 20&WR&IR&CWR &WR&WR&QWR\\
 30&WR&IR&CWR &WR&WR&WR\\
 40&WR&CWR&CWR &WR&WR&WR\\
 50&WR&CWR&CWR &WR&WR&WR\\
 60&WR&QWR&CWR &WR&WR&WR\\
 70&WR&QWR&CWR &WR&WR&WR\\
 80&WR&QWR&CWR &WR&WR&WR\\
 90&WR&QWR&CWR &WR&WR&WR\\
 100&WR&QWR&CWR &Rolls&WR&WR\\
 120&WR&WR&CWR &Rolls&WR&WR\\
\end{tabular}
\end{ruledtabular}
\end{table}

\begin{figure}[ht]
\begin{center}
\resizebox{0.45\textwidth}{!}{%
  \includegraphics{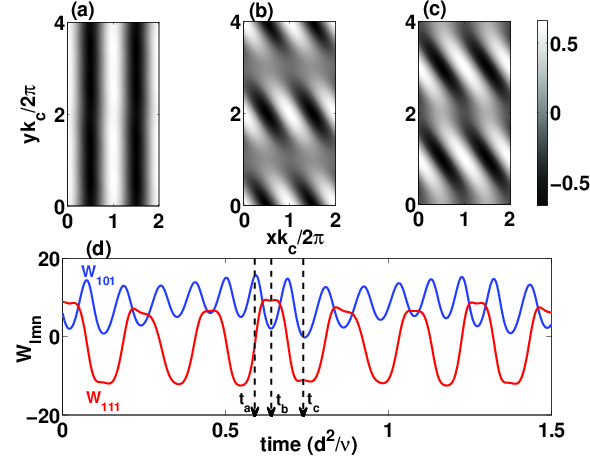}
}
\caption{\label{fig:half_cont}(Color online) Mid-plane ($z=1/2$) contour plots of the convective temperature field for $\mathrm{Pr}=0.025$ and $r = 1.1$ in a simulation box of rectangular cross-section ($\eta = 2$). A quasiperiodic competition between straight and inclined wavy rolls for $\mathrm{Q} = 60$ [$k_\mathrm{c} (\mathrm{Q}=60) = 3.376$] shown for three different instants. (d) Temporal variations of the Fourier modes $W_{101}$ [blue (black) curve] and $W_{111}$ [red (gray) curve]. The time instants marked  by $t_a$, $t_b$ and $t_c$ correspond to the patterns (a), (b) and (c) respectively.}
\end{center}
\end{figure}

\begin{figure}[ht]
\begin{center}
\resizebox{0.45\textwidth}{!}{%
  \includegraphics{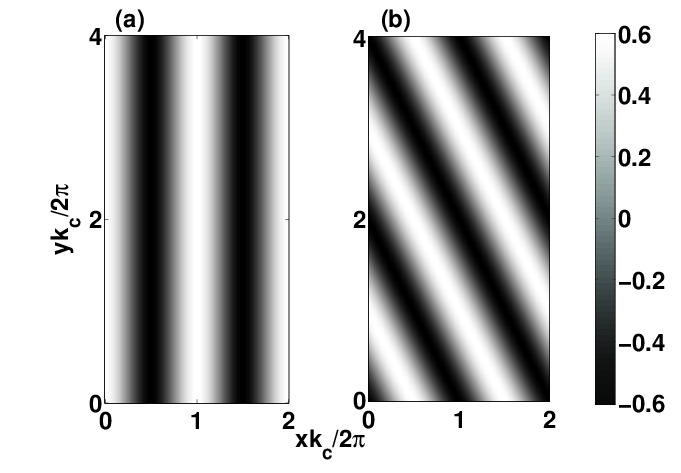}
}
\caption{\label{fig:half_stat} Contour plots of the convective temperature field (at $z=0.5$) in a rectangular simulation box ($\eta = 2$) for $\mathrm{Pr} = 0.1$ showing (a) stationary straight rolls  for $r = 1.05$ and $\mathrm{Q} = 100$ [$k_\mathrm{c} (\mathrm{Q}=100) =3.702$] and (b) stationary inclined rolls (IR) for  $r = 1.2$ and $\mathrm{Q} = 10$ [$k_\mathrm{c} (\mathrm{Q}=10) = 2.589$].}
\end{center}
\end{figure}

As the reduced Rayleigh number is further raised, we observe interesting patterns. The longer simulation box allows the turning of rolls, which is not observed in a small square simulation box. The turning of rolls has a similarity with pattern dynamics due to  K\"{u}ppers-Lortz instability~\cite{kl_1969} in thermal convection in the presence of Coriolis force. The Coriolis force excites the vertical vorticity in rotating convection. Large wavelength perturbations in low-Prandtl-number fluids allow the generation of vertical vorticity easily even in the absence of rotation.  The wavy rolls orient themselves  in the horizontal plane making an angle from its original position. The temporal behavior is either chaotic or quasiperiodic. The upper row of Fig.~\ref{fig:half_cont} show the mid-plane ($z=1/2$) contour plots for the temperature field at three different instants for  $\mathrm{Pr}=0.025$, $\mathrm{Q}=60$ and $r=1.1$. The lower row of Fig.~\ref{fig:half_cont} displays the variation of the two leading modes with time for these patterns. The fluid patterns shown in  Fig.~\ref{fig:half_cont}(a), (b) and (c) are for the instants marked by $t_a$, $t_b$ and $t_c$ respectively in fig.~\ref{fig:half_cont} (d). The patterns show the appearance of straight rolls and wavy rolls oriented at an angle with the straight rolls. The oriented wavy rolls appear when the magnitude of the nonlinear mode $W_{111}$ is much larger than that of the roll mode $W_{101}$. The leading modes corresponding to these patterns vary quasiperiodically in time. We also observe stationary oblique (inclined) rolls (IR) at smaller values of $\mathrm{Q}$. Figures~\ref{fig:half_stat}(a) and \ref{fig:half_stat}(b) show the mid-plane contour plots of 2D stationary rolls close to the onset of convection for $\mathrm{Pr}=0.1$, $\mathrm{Q}=100$, $r=1.05$ and stationary inclined rolls at tertiary instability for $\mathrm{Pr}=0.1$, $\mathrm{Q}=10$, $r=1.2$. 

\section{Conclusions}
We have investigated the effect of a uniform vertical magnetic field on the Rayleigh-B\'{e}nard convection for zero-magnetic-Prandtl-number fluids considering stress-free top and bottom surfaces using direct numerical simulations. The magnetic field strongly delays the primary, secondary and higher order instabilities. Convection appears in the form of stationary straight rolls at the primary instability. The Nusselt number scales with the relative distance from the instability onset $\epsilon$ as $\epsilon^{0.91}$ close to onset of convection. The Nusselt number also scales with $\mathrm{Ra/Q}$ as $\mathrm{Nu\sim (Ra/Q)}^{\mu}$ for $\mathrm{Ra/Q} > 25$. The scaling exponent $\mu$ depends on $\mathrm{Q}$, and its values agree with experimental results.  The straight rolls become wavy at the secondary instability showing  periodic, quasiperiodic or chaotic behavior in time in a square simulation box for different values of $\mathrm{Pr}$, $r$ and $\mathrm{Q}$.  The ratio $r_{\circ}/\mathrm{Pr}$ is independent of $\mathrm{Q}$, while $\mathrm{[Ra_{\circ} (Q) - Ra_{\circ} (Q=0)]/Pr}$ varies linearly with $\mathrm{Q}$ for smaller values of $\mathrm{Q}$. In a rectangular simulation box, oblique wavy rolls as well as oblique stationary rolls are observed close to onset of convection.  

\noindent{\bf Acknowledgements:} We have benefited from fruitful discussions with Pinaki Pal, Priyanka Maity and Hirdesh Pharasi. 


\begin{thebibliography}{99}
\bibitem{chandra}
{S. Chandrasekhar}, \textit{Hydrodynamic and Hydromagnetic Stability}, {Oxford University Press, London} (1961).
\bibitem{nakagawa57}
{Y. Nakagawa}, {Proc. R. Soc. Lond. A} {\bf 240}, 108 (1957). 
\bibitem{nakagawa59}
{Y. Nakagawa}, {Proc. R. Soc. Lond. A} {\bf 249}, 138 (1959).
\bibitem{proctor_weiss_1982}
{M. R. E. Proctor and N. O. Weiss}, {Rep. Prog. Phys.} {\bf 45}, 1317 (1982).
\bibitem{glatzmaier_etal_1999}
G. Glatzmaier, R. Coe, L. Hongre, and P. Roberts, Nature (London) {\bf 401}, 885 (1999).
\bibitem{cattaneo_etal_2003}
F. Cattaneo, T. Emonet, and N. Weiss, Astrophys. J. {\bf 588}, 1183 (2003).
\bibitem{rucklidge06}
{A. M. Rucklidge, M. R. E. Proctor and J. Prat}, {Geo. Astr. Fluid Dyn.} {\bf 100}, 121 (2006).
\bibitem{fauve_etal_1981}
{S. Fauve, C. Laroche and A. Libchaber}, {J. Phys. Lett.} {\bf 42}, L455 (1981).
\bibitem{knobloch_etal_1981}
{E. Knobloch, N. O. Weiss and L. N. Da Costa}, {J. Fluid Mech.} {\bf 113}, 153 (1981). 
\bibitem{bc82}
{F. H. Busse and R. M. Clever}, {Phys. Fluids} {\bf 25}, 931 (1982).
\bibitem{fauve_etal_1984}
S. Fauve, C. Laroche, A. Libchaber, and B. Perrin, Phys. Rev. Lett. {\bf 52}, 1774 (1984).
\bibitem{meneguzzi87}
{M. Meneguzzi, C. Sulem, P. L. Sulem and O. Thual}, {J. Fluid Mech.} {\bf 182}, 169 (1987).
\bibitem{cb89}
{R. M. Clever and F. H. Busse}, {J. Fluid Mech.} {\bf 201}, 507 (1989).
\bibitem{houchens_2002}
{B. C. Houchens, L. M. Witkowski and J. S. Walker}, {J. Fluid Mech.} {\bf 469}, 189 (2002).
\bibitem{dawes_2007}
{J. H. P. Dawes}, {J. Fluid Mech.} {\bf 570}, 385 (2007).
\bibitem{podvigina_2010}
{O. Podvigina}, {Phys. Rev. E} {\bf 81}, 056322 (2010).
\bibitem{pk_2012}
{P. Pal and K. Kumar}, {Eur. Phys. J. B} {\bf 85}, 201 (2012).
\bibitem{cioni_etal_2000}
{S. Cioni, S. Chaumat and J. Sommeria}, {Phys. Rev. E} {\bf 62}, R4520 (2000).
\bibitem{ao_2001}
{J. M. Aurnou and P. L. Olson}, {J. Fluid Mech.} {\bf 430}, 283 (2001).
\bibitem{bm_2002}
{U. Burr and U. M{\"{u}}ller}, {J. Fluid Mech.} {\bf 453}, 345 (2002).
\bibitem{yanagisawa_etal_2011}
{T. Yanagisawa, Y. Yamagishi, Y. Hamano, Y. Tasaka, and Y. Takeda}, {Phys. Rev. E} {\bf 83}, 036307 (2011).
\bibitem{goldstein_graham_1969}
{R.J. Goldstein and D.J. Graham}, {Phys. Fluids.} {\bf 12}, 1133 (1969).
\bibitem{lee_etal_2008}
{B. Lee, J. Z. Liu, B. Sun, C. Y. Shen and G. C. Dai}, {Express Polymer Lett.} {\bf 2}, 357 (2008).
\bibitem{kl_1969}
{G. K\"{u}ppers  and D. Lortz}, {J. Fluid. Mech.} {\bf 35}, 609 (1969).

\end{thebibliography}
\end{document}